\documentclass{article}
\usepackage{amsmath}
\usepackage{cite}
\usepackage{graphicx}
\usepackage{dcolumn}

\begin{document}

\date{}
\title{On a recent calculation of the mass spectra of quarkonia using the Cornell
potential with spin-spin interactions}
\author{Francisco M. Fern\'{a}ndez \\
INIFTA, Divisi\'on Qu\'imica Te\'orica,\\
Blvd. 113 (S/N), Sucursal 4, Casilla de Correo 16, \\
1900 La Plata, Argentina}
\maketitle

\begin{abstract}
We analyse a recent application of the Cornell potential with spin-spin
interaction to the mass spectra of quarkonia and show that the authors have
in fact used the Kratzer-Fues potential. They inadvertently converted one
potential into the other by means of an invalid transformation.
\end{abstract}

In a paper published recently, Omugbe et al\cite{OEA23} tried to it
experimental data by means of the solutions of the non-relativistic Schr\"{o}%
dinger equation with the Cornell potential with spin-spin interaction. They
carried out a coordinate transformation and two approximations that enabled
them to derive an eigenvalue equation that can be solved by the
Nikiforov-Uvarov method. In this way they obtained analytical approximate
expressions for the eigenvalues and eigenfunctions. In this short note we
analyse the mathematical procedure followed by those authors.

In order to study a meson mass spectra Omugbe et al\cite{OEA23} proposed the
Cornell potential with spin-spin interaction
\begin{eqnarray}
V(r) &=&-\frac{4\alpha _{s}}{3r}+br+C_{s}e^{-\sigma ^{2}r^{2}},  \nonumber \\
C_{s} &=&16\alpha _{s}\pi \left( \frac{\sigma }{\sqrt{\pi }}\right) ^{3}%
\frac{s(s+1)-\frac{3}{2}}{9m_{q}m_{\bar{q}}},  \label{eq:V(r)}
\end{eqnarray}
where $\alpha _{s}$, $b$ and $\sigma $ are adjustable model parameters, $s$
is the spin quantum number and $m_{q}$, $m_{\bar{q}}$ are the masses of the
quark and antiquark, respectively. They focused on the radial part of the
eigenvalue equation written as
\begin{equation}
-\frac{\hbar ^{2}}{2\mu }\frac{d^{2}\psi _{nl}(r)}{dr^{2}}+\left[ V(r)+\frac{%
\hbar ^{2}l(l+1)}{2\mu r^{2}}\right] \psi _{nl}(r)=E_{nl}\psi
_{nl}(r), \label{eq:Schrodinger}
\end{equation}
where $\mu $ is the reduced mass and $l$ the rotational quantum number. The
authors did not mention it explicitly but we assume that the boundary
conditions are $\psi _{nl}(0)=0$ and $\psi _{nl}(r\rightarrow \infty )=0$
provided that $b>0$.

Since the eigenvalue equation (\ref{eq:Schrodinger}) is not
exactly solvable, Omugbe er al\cite{OEA23} resorted to a series of
approximations. The first one consists of expanding the Gaussian
term to second order which produces the approximate potential
\begin{equation}
V^{OEA}(r)=C_{s}-\frac{4\alpha _{s}}{3r}+br-C_{s}\sigma ^{2}r^{2}.
\label{eq:V^(OEA)(r)}
\end{equation}
For $s=0$ this potential supports bound states because $C_{0}<0$; however,
for $s=1$ there are no bound states because $C_{1}>0$ and the resulting
potential is unbounded from below. In other words: this approximation is
unsuitable for triplet states. Omugbe et al\cite{OEA23} overlooked this
limitation that was overcome by the second incorrect approximation discussed
below.

\begin{figure}[tbp]
\begin{center}
\includegraphics[width=9cm]{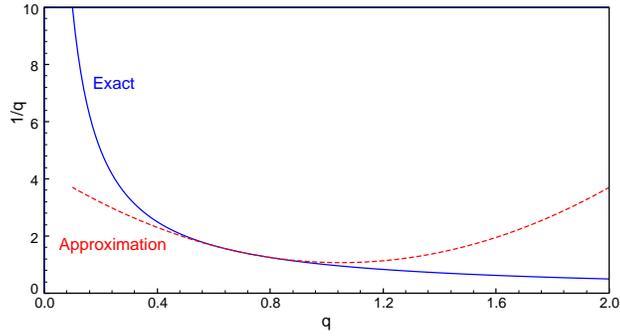}
\end{center}
\caption{Exact values of $1/q$ (blue, continuous line) and the approximation
(\ref{eq:Second_approx}) (red, dashed line) for $\delta =0.7$}
\label{Fig:1q}
\end{figure}

\begin{figure}[tbp]
\begin{center}
\includegraphics[width=9cm]{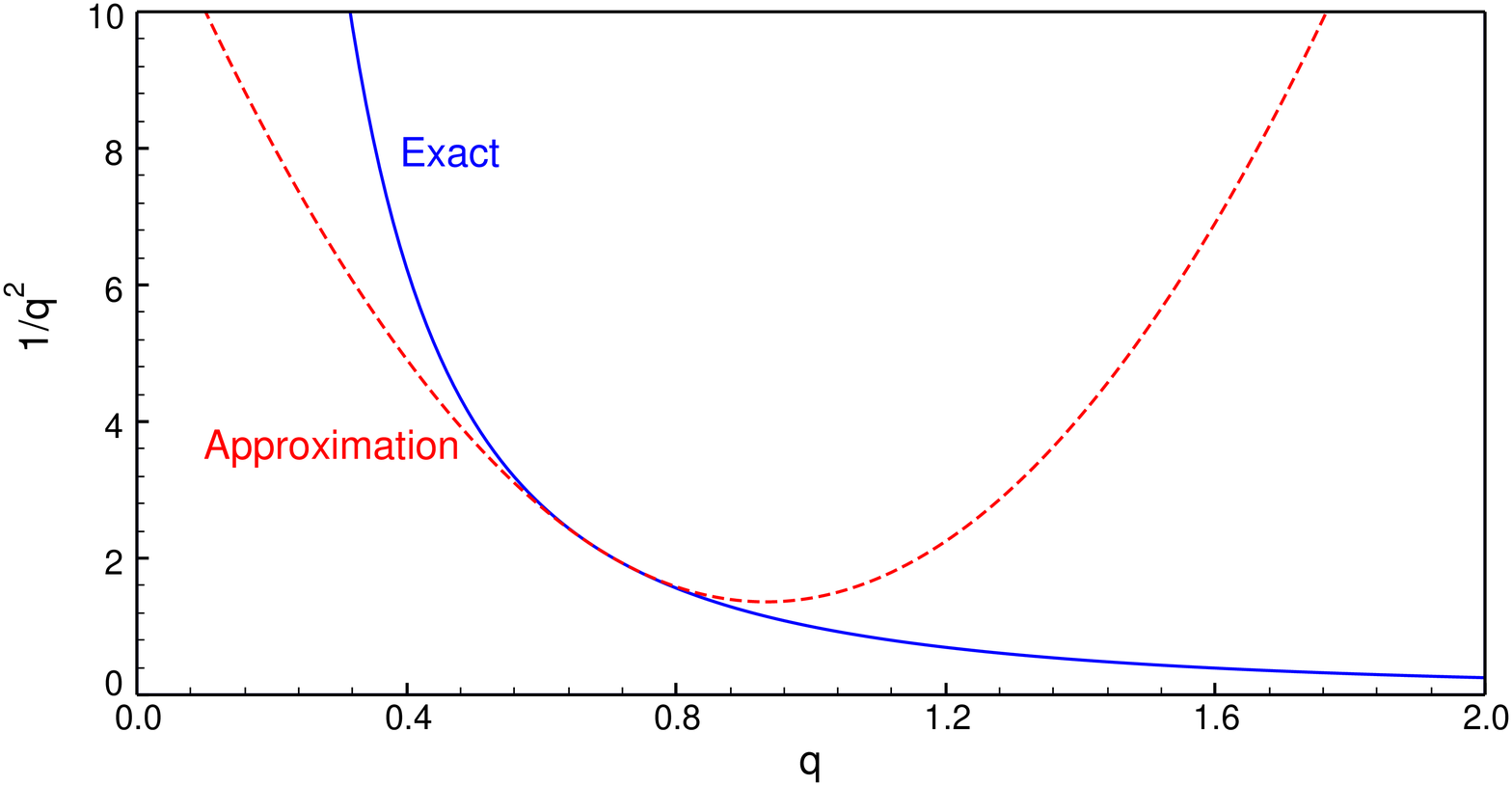}
\end{center}
\caption{Exact values of $1/q^2$ (blue, continuous line) and the
approximation (\ref{eq:Second_approx}) (red, dashed line) for $\delta =0.7$}
\label{Fig:1q2}
\end{figure}

Omugbe et al\cite{OEA23} carried out the variable transformation $q=1/r$
that leads to
\begin{equation}
\frac{d^{2}\psi }{dq^{2}}+\frac{2}{q}\frac{d\psi }{dq}+\frac{1}{q^{4}}\left(
\epsilon _{nl}+aq-\frac{c}{q}+\frac{d}{q^{2}}-Lq^{2}\right) \psi =0,
\label{eq:diff_eq_q_1}
\end{equation}
where all the parameters are given in their paper. Note that we have taken
into consideration that $2q/q^{2}=2/q$. At this point, Omugbe et al\cite
{OEA23} argued as follows: ``However, we must put the equation into a
standard form using an approximation scheme by expanding $c/q$ and $d/q^{2}$
in power series to second-order around $q_{0}$ ($\delta =1/q_{0}$) which is
assumed to be the characteristic radius of mesons.'' From this second
approximation they obtained the highly inaccurate expressions
\begin{eqnarray}
\frac{c}{q} &=&c\left( \frac{3}{\delta }-\frac{3q}{\delta ^{2}}+\frac{q^{2}}{%
\delta ^{3}}\right) ,  \nonumber \\
\frac{d}{q^{2}} &=&d\left( \frac{6}{\delta ^{2}}-\frac{8q}{\delta ^{3}}+%
\frac{3q^{2}}{\delta ^{4}}\right) .  \label{eq:Second_approx}
\end{eqnarray}
Since, for each of these equations, the left-hand side is completely
different from the right-hand side, as shown in figures~\ref{Fig:1q} and \ref
{Fig:1q2} for $\delta =0.7$, we conclude that Omugbe et al\cite{OEA23}
transformed the original model into another one with completely different
behaviour at $q\rightarrow 0$ and $q\rightarrow \infty $, given by the
eigenvalue equation
\begin{equation}
\frac{d^{2}\psi }{dq^{2}}+\frac{2}{q}\frac{d\psi }{dq}+\frac{1}{q^{4}}\left(
-Aq^{2}+Bq-C\right) \psi =0,  \label{eq:diff_eq_q_2}
\end{equation}
where all the parameters are given in their paper. Note that, once more, we
have taken into account that $2q/q^{2}=2/q$.

The eigenvalue equation (\ref{eq:diff_eq_q_2}) describes a physical model
that is completely different from the original one given by $V(r)$ (\ref
{eq:V(r)}) and even from the approximate one based on $V^{OEA}(r)$ (\ref
{eq:V^(OEA)(r)}). In fact, if we substitute $q=1/r$ into equation (\ref
{eq:diff_eq_q_2}) we obtain the eigenvalue equation
\begin{equation}
\frac{d^{2}\psi }{dr^{2}}+\left( \frac{B}{r}-\frac{A}{r^{2}}-C\right) \psi
=0,  \label{eq:diff_eq_r_OEA}
\end{equation}
where $V^{KF}(r)=\frac{\hbar ^{2}}{2\mu }\left( -\frac{B}{r}+\frac{A}{r^{r}}%
\right) $ resembles the Kratzer-Fues potential\cite{K20,F26}. The eigenvalue
equation (\ref{eq:diff_eq_r_OEA}) can be easily solved by means of the
simple and straightforward Frobenius (power-series) method\cite{F21}. Omugbe
et al\cite{OEA23} applied the far more complicated Nikiforov-Uvarov method
to equation (\ref{eq:diff_eq_q_2}) and obtained the eigenvalues
\begin{equation}
E_{nl}^{OEA}=\Gamma _{0}-\frac{\mu }{2\hbar ^{2}}\left( \frac{\Gamma _{1}}{%
n+\Gamma _{2}}\right) ^{2},  \label{eq:E^(OEA)}
\end{equation}
where the parameters $\Gamma _{j}$ are given in their paper. As expected,
this spectrum resembles the one for the Kratzer-Fues potential\cite{F26,F21}.

It is illustrative to compare the spectrum of the original model based on
the potential-energy function (\ref{eq:V(r)}) and the one used by Omugbe et
al\cite{OEA23} in their study of the physical problem given by either
equation (\ref{eq:diff_eq_r_OEA}) or (\ref{eq:diff_eq_q_2}). Since the
original model potential is unbounded from above (increases as $br$ when $%
r\rightarrow \infty $), we conclude that $E_{nl}\rightarrow \infty $ when $%
n\rightarrow \infty $. On the other hand, $E_{nl}^{OEA}\rightarrow \Gamma
_{0}$ when $n\rightarrow \infty $. The reason for this gross qualitative
discrepancy is due to the transformations (\ref{eq:Second_approx}) that
convert the original model into a completely different one.

\textbf{Summarizing}: Omugbe et al\cite{OEA23} claimed that they fitted
experimental data to the solutions of the Schr\"{o}dinger equation for the
Cornell potential with spin-spin interaction when the true fact is that they
used the Kratzer-Fues potential that exhibits a completely different
spectrum. The origin of this mistake can be traced back to the two invalid
transformations discussed above.

\end{document}